\documentclass[conference]{IEEEtran}

\IEEEoverridecommandlockouts

\usepackage{cite}
\usepackage{amsmath,amssymb,amsfonts}
\usepackage{algorithmic}
\usepackage{graphicx}
\usepackage{textcomp}
\usepackage{xcolor}
\usepackage{flushend}
\usepackage{siunitx}
\def\BibTeX{{\rm B\kern-.05em{\sc i\kern-.025em b}\kern-.08em
    T\kern-.1667em\lower.7ex\hbox{E}\kern-.125emX}}

\begin{document}

\title{Spectral Characterization of a 90 GHz CLASS Pixel}

\author{
\IEEEauthorblockN{Gregory Jaehnig\textsuperscript{1,2},
John Appel\textsuperscript{3},
Sarah Marie Bruno\textsuperscript{3},
Jake Connors\textsuperscript{4},
Shannon M. Duff\textsuperscript{2},\\ 
Naina Gupta\textsuperscript{3},
Johannes Hubmayr\textsuperscript{2},
Matthew A. Koc\textsuperscript{1,2},
Tammy Lucas\textsuperscript{2},\\ 
Tobias Marriage\textsuperscript{3},
Lola Morales Perez\textsuperscript{3},
Caleigh Ryan\textsuperscript{3},
Jeff Van Lanen\textsuperscript{2}
\thanks{We acknowledge the National Science Foundation Division of Astronomical Sciences for their support of CLASS under grant Nos. 0959349, 1429236, 1636634, 1654494, 2034400, 2102694, 2109311, and 2442928. Detector development work at JHU was funded by NASA cooperative agreement 80NSSC19M0005.}
}
\IEEEauthorblockA{\textsuperscript{1}Department of Physics, University of Colorado, Boulder, CO}
\IEEEauthorblockA{\textsuperscript{2}Quantum Sensors Division, National Institute of Standards and Technology, Boulder, CO}
\IEEEauthorblockA{\textsuperscript{3}Department of Physics \& Astronomy, Johns Hopkins University, Baltimore, MD}
\IEEEauthorblockA{\textsuperscript{4}NASA Goddard Space Flight Center, Greenbelt, MD}
}

\maketitle

\begin{abstract}
The Cosmology Large Angular Scale Surveyor (CLASS) is an experiment designed to measure the polarization of the cosmic microwave background on large angular scales to probe cosmic reionization and search for the inflationary $B$-mode signal.
CLASS is a multi-frequency ensemble of telescopes with bands centered at 40, 90, 150, and \SI{220}{GHz}.
Each telescope has arrays of feedhorn-coupled transition edge sensor bolometers at the focal plane.
The frequency response is primarily defined by the on-chip bandpass filter with additional contributions coming from the feedhorn, orthomode transducer, and 180-degree hybrid.
In this study, we compare simulations and measurements of the frequency response of single pixel witness devices in the \SI{90}{GHz} band with and without the bandpass filter.
For the first time, we can separate the effects of the bandpass filter from the other microwave components using Fourier transform spectroscopy and design splits of the pixel.
The results show that the $\mathbf{\text{\SI[bracket-negative-numbers = false]{-3}{dB}}}$ band edges are at \SI{80}{GHz} and \SI{108}{GHz}.
The measurements demonstrate a robust method for characterizing the spectral response of individual components, which is crucial for optimizing the performance of future detector arrays.
\end{abstract}

\begin{IEEEkeywords}
Transition Edge Sensor, On-Chip Filters, Fourier Transform Spectroscopy, Cosmic Microwave Background, Polarimeter
\end{IEEEkeywords}

\section{Introduction}

The standard model of cosmology has been integral in understanding the composition and evolution of the universe.
However, it leaves open the question of the origin of density perturbations that seeded large-scale structure~\cite{lyth2009primordial}.
An extension of the model posits an early inflationary stage as the source of perturbations as well as a resolution to the horizon and flatness problems~\cite{guth1981inflationary}.
Testing theories of inflation requires a new generation of experiments with unprecedented sensitivity and control of systematics.

Cosmology Large Angular Scale Surveyor (CLASS) is optimized to measure the polarization of the cosmic microwave background (CMB) over large angular scales, from 1 to 90 degrees, where a primordial $B$-mode signal could dominate over the lensing signal.
This angular range captures both reionization and recombination peaks, a simultaneous detection of which would bolster confidence in the theory of inflation~\cite{essinger2014class}.
CLASS is also well suited to measure the optical depth to reionization using $E$ modes, a measurement that requires observations from $\ell < 20$~\cite{li2025measurement}.
Galactic foreground emissions can be removed by measuring the signal in multiple frequency bands to separate the falling spectrum of synchrotron and rising spectrum of dust from the blackbody spectrum of the CMB.

The detector of choice to accomplish this is the transition edge sensor (TES) bolometer, which can achieve background-limited performance for all bands~\cite{appel2019sky}.
The second CLASS \SI{90}{GHz} telescope will receive four new detector arrays in 2025.
This paper covers witness pixels that were fabricated alongside these new arrays.

The optimal frequency response for achieving background-limited performance in a ground-based CMB experiment is one that avoids atmospheric emission lines, while maximizing bandwidth within the atmospheric window.
For the \SI{90}{GHz} band, there are a set of strong oxygen absorption lines at \SI{60}{GHz} and a set of water and oxygen absorption lines at \SI{118}{GHz}~\cite{fox2024evaluation}.
If these lines couple to the detector the loading on them will increase, raising noise levels, while not increasing sensitivity to the CMB.
Before fielding an instrument, it is crucial to verify these conditions in the laboratory.
One difficulty in interpreting the measurements is how the setup itself can affect the response through standing waves, filters, and misaligned optics.
The measured spectral properties $(I)$ of a linear system can be written as a product of several transfer functions in frequency space: the on-chip bandpass filter $(B)$, the on-chip low-pass filter $(L)$, the combined on-chip filter response $(BL)$, the quasi-optical low-pass filter $(Q)$, the neutral density filter $(N)$, and all other spectral effects in $(F)$.
It can be shown that, with impedance-matched components, taking ratios of spectra where one factor is removed allows us to isolate individual terms~\cite{oppenheim1997signals_ch4sec5}.
We will do this for the terms $BL$, $Q$, and $N$.
The ratio method will also reduce the impacts of some non-idealities in the measurement setup because they will be mostly common mode.
For example, standing waves between the source and detector create fringes in the measured spectrum, but ought to be common to both measurements and cancel.

Next, we will examine the design of the pixel, followed by the experimental setup, and finish by presenting the results of each measurement run and discussing them.

\section{Pixel Design}

\begin{figure}[htbp]
\centerline{\includegraphics[width=\columnwidth]{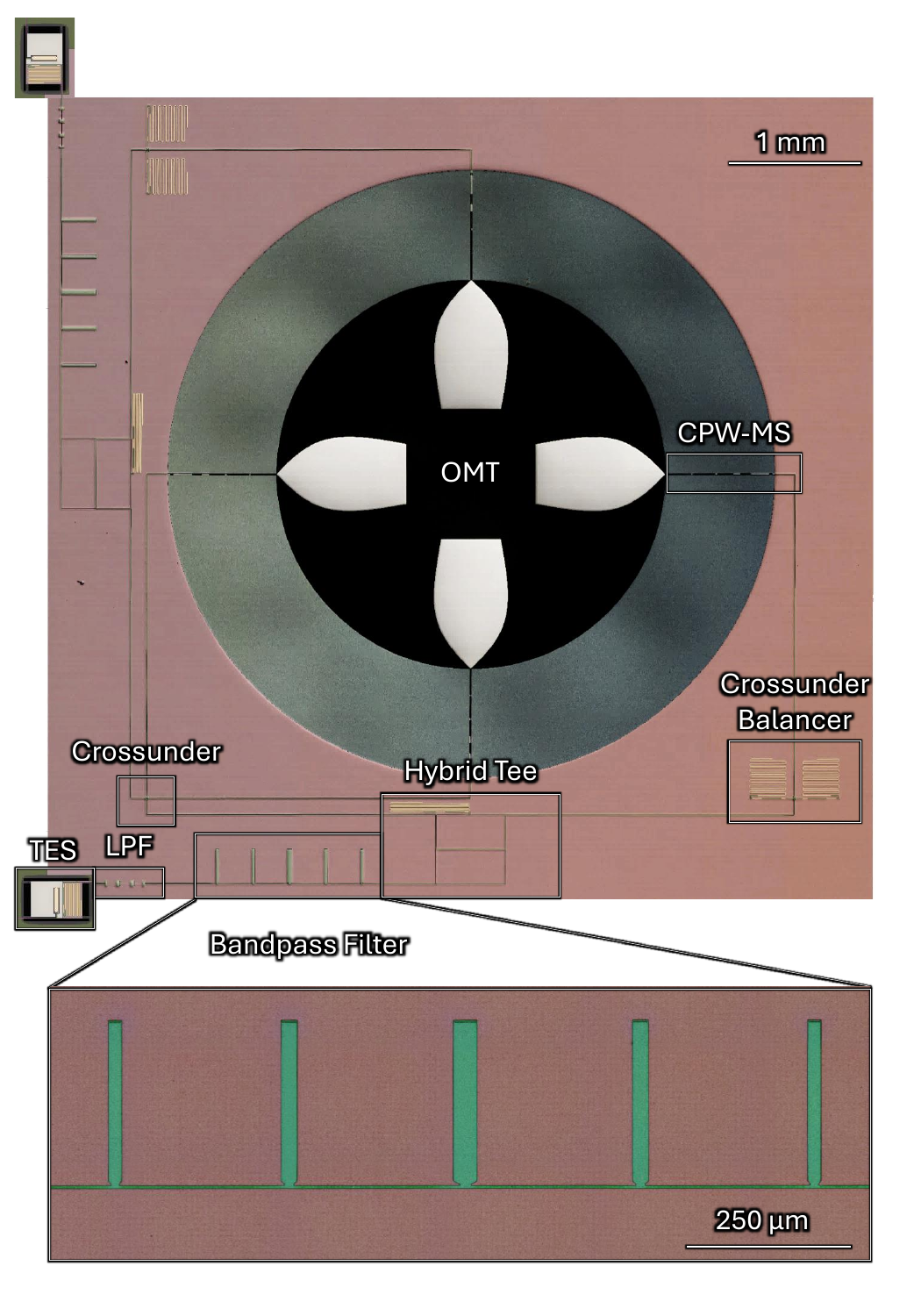}}
\caption{Confocal microscope image of the pixel from a 3D optical profilometer with mm-wave components labeled.
The mm-wave components for the orthogonal polarization can be seen on the upper left side of the image.
Images of the TESs under higher magnification are inset in the corners.
An image of the bandpass filter is shown on the bottom with the microstrip running horizontally and the five stubs extending above.}
\label{3Dpixel}
\end{figure}

\begin{figure}[htbp]
\centerline{\includegraphics[width=\columnwidth]{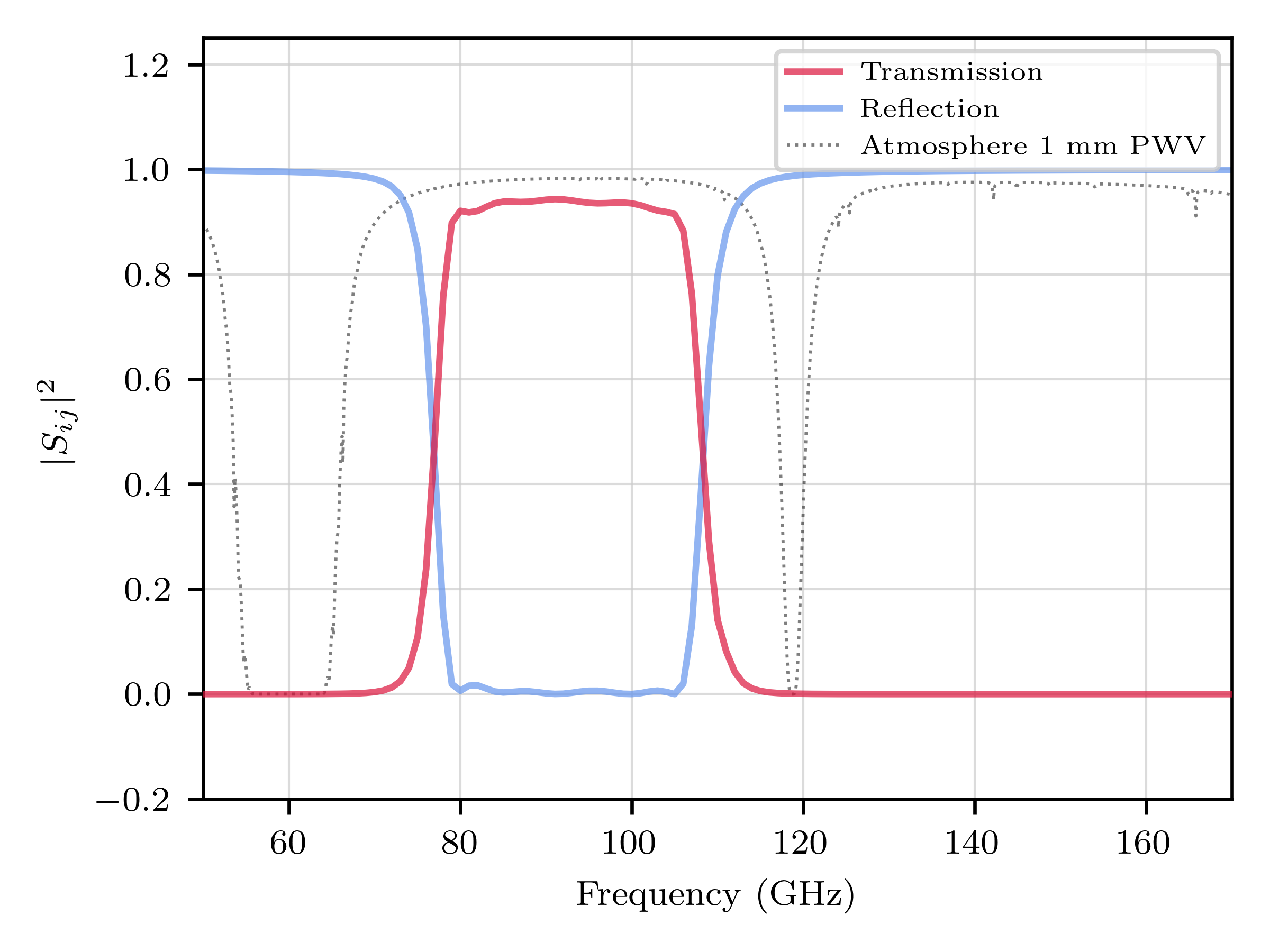}}
\caption{Simulated transmission and reflection of the band pass filter design. The atmospheric transmission with 1 mm of precipitable water vapor is shown as the dotted line. The passband falls squarely within the atmospheric transmission window~\cite{pardo2002atmospheric}.}
\label{simulation}
\end{figure}

To understand the pixel design we can follow the light from the sky to the detector.
The components of the pixel are shown with labels in Figure~\ref{3Dpixel}.
Light enters a feedhorn that transitions to a circular waveguide above a niobium orthomode transducer (OMT) suspended on a \SI{2}{\um} thick silicon nitride (SiN) membrane.
The OMT has a diameter optimized for waveguide with a cutoff frequency, $f_c$, of \SI{60}{GHz} and an upper edge with $\mathbf{\text{\SI[bracket-negative-numbers = false]{-3}{dB}}}$ co-polar coupling is ${\sim}2.2f_c$ at \SI{132}{GHz}~\cite{hubmayr2022tolerance}.
The high impedance OMT is matched to co-planar waveguide (CPW) and transitions to low impedance microstrip (MS) lines.
The MS lines carry the wave from the orthogonal polarizations and cross one another before being combined in a hybrid tee where unwanted sum modes are terminated.
The MS continues to the on-chip filters: a 5-pole quarter-wave shorted-stub bandpass filter (BPF) followed by a stepped impedance lowpass filter (LPF) to reduce the $3f$ harmonic of the BPF.
A simulation of the BPF is shown in Figure~\ref{simulation}.
The MS terminates on a lossy gold meander adjacent to an AlMn TES on a SiN island.
A PdAu thermal ballast limits the detector bandwidth appropriate to the time-domain multiplexing system used to readout the TES signal~\cite{yoon2009feedhorn,Lanting2005}.
This design simplifies previous ones in that a single set of on-chip filters are placed after the hybrid tee rather than two sets before it~\cite{hubmayr2015feedhorn}.
We measured the deformation of the OMT membrane with a 3D optical profilometer and found a peak-to-peak height variation of \SI{20}{\um} over a \SI{4.7}{\mm} diameter which is negligible compared to one wavelength.

\section{Experimental Setup}

The witness pixels were tested side by side in a cryostat with a 2-stage adiabatic demagnetization refrigerator (ADR) backed by a pulse tube cooler.
The optical path of the cryostat consists of a room temperature closed-cell foam vacuum window, a \SI{20}{\mm} thick polytetrafluoroethylene (PTFE) filter at \SI{50}{K} to block IR, and a set of interchangeable filters at \SI{4}{K} to further block IR.
We have the option of using a \SI{4}{K} neutral density filter (NDF) to reduce in-band loading on the detector or another PTFE filter.
The NDF is a \SI{16}{mm} thick slab of Eccosorb MF-110, a magnetically loaded machinable epoxy~\cite{Laird_ECCOSORB_MF}.
There are \SI{500}{\um} thick etched teflon anti-reflective coatings on both sides.
We use quasi-optical low pass filters to limit the IR loading with cutoffs at \SI{5.8}{\per\cm}, \SI{10}{\per\cm}, and \SI{11}{\per\cm}~\cite{Ade2006filter}.
This setup uses a smooth-walled metal feedhorn with a circular waveguide radius of \SI{1.2}{mm} and cutoff frequency of \SI{73}{GHz}.

\begin{figure}[htbp]
\centerline{\includegraphics[width=1\columnwidth]{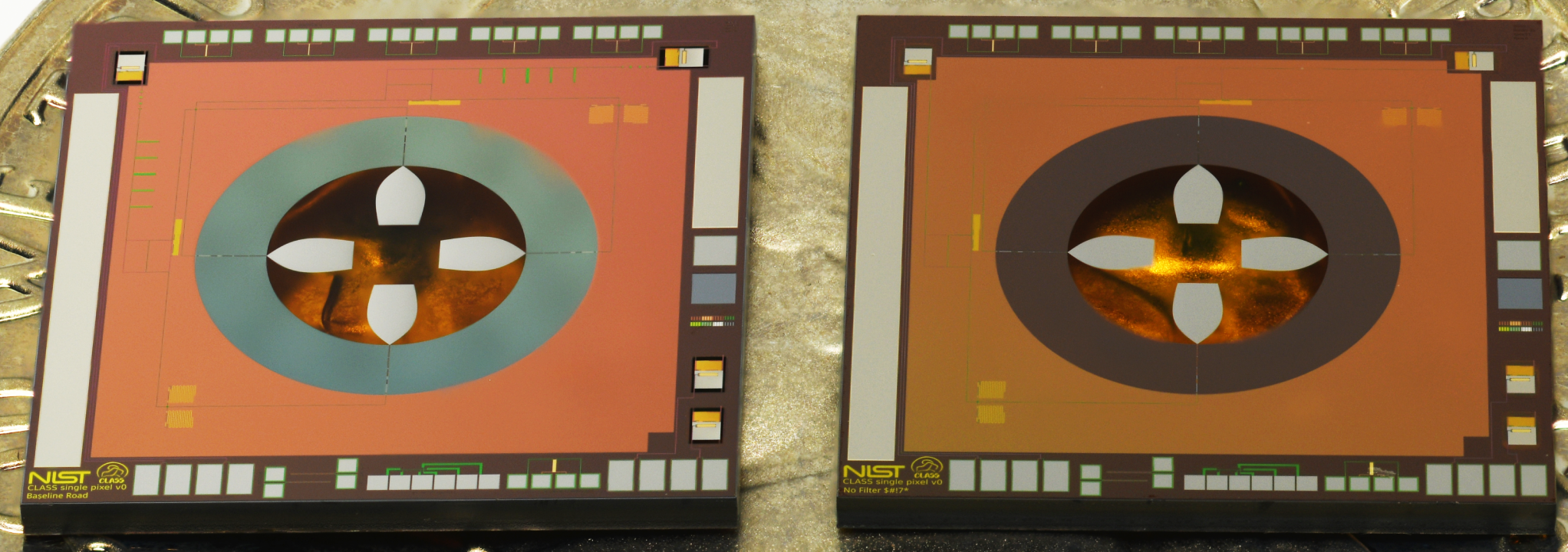}}
\caption{Photograph of two 6$\times$\SI{6}{\mm} witness pixels from the first of five fabricated arrays. The pixel on the left has on-chip bandpass and lowpass filters, while the pixel on the right has these features replaced by a run of microstrip transmission line.}
\label{2pixels}
\end{figure}

The TESs were biased near the middle of their superconducting transition and read out with a time-division multiplexed readout system~\cite{irwin2005transition,irwin2002time}.
The spectra were measured with a Martin–Puplett interferometer that uses a series of wire-grid polarizers to split and recombine the differentially polarized light from a room temperature and a 1050~$^{\circ}$C blackbody, with one beam reflecting off a fixed rooftop mirror and the other off a moving rooftop mirror to create an interference pattern that is the Fourier transform of the detector spectrum~\cite{martin1970polarised,martin1982infrared}.
The output polarizer of the interferometer was oriented 45 degrees from the polarization axes of the OMTs to obtain data from both polarizations simultaneously while incurring a factor of $\sqrt{2}$ penalty in the detected signal.

In the data analysis, we Fourier transform the interferogram in real space to the spectrum in frequency space.
We notch out frequencies related to the AC line power and its harmonics.
Then every scan is flattened by a 1st order polynomial and folded about the point of zero path length difference to create a double-sided interferogram.
The phase correction is applied directly to the interferogram following Richards' method~\cite{richards1964high}.
Each measurement is an average of 100 scans, which results in a noise level of less than 0.5\% of the normalized amplitude.
The transmission, in arbitrary units, is normalized by fitting to a model of the bandpass filter, which results in portions of the spectrum exceeding unity transmission.
The frequency resolution of these measurements is \SI{0.5}{GHz}.

\section{Results}
We tested two pixels, shown in Figure~\ref{2pixels}, over the course of three runs with different 4~K optical filters in the cryostat each time.
The first measurement run used \SI{10}{\per\cm} and \SI{11}{\per\cm} LPFs with an NDF, the second run replaced the \SI{10}{\per\cm} and \SI{11}{\per\cm} LPFs with a \SI{5.8}{\per\cm} LPF and replaced the NDF with a PTFE filter, and the third continued to use the \SI{5.8}{\per\cm} LPF but replaced the PTFE filter with the NDF.
The final run has the most complete data set, but the first two runs still have useful data and can be used with the ratio method to isolate the frequency response of individual components.

\subsection{Measurement Run 1}

\begin{figure}[htbp]
\centerline{\includegraphics[width=\columnwidth]{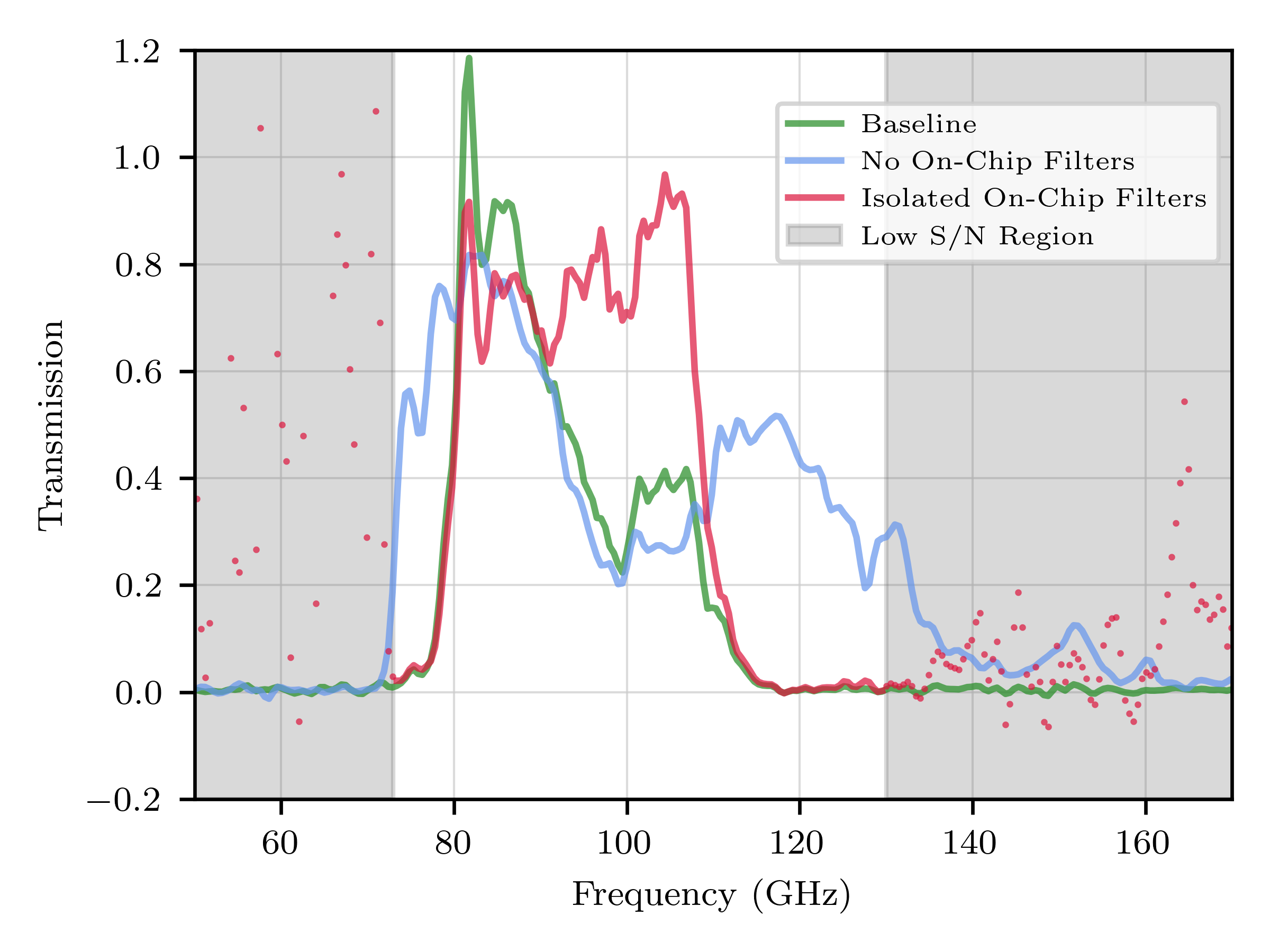}}
\caption{Spectra from both baseline and no on-chip filter pixels from run~1. The \SI{10}{\per\cm} and \SI{11}{\per\cm} quasi-optical low pass filters have distorted the spectrum near \SI{100}{GHz}.
The ratio of each spectra yield the isolated on-chip filter response which has a much flatter passband in comparison. The ratio fluctuates outside the band edges due to low signal-to-noise ratio in the no on-chip filter pixel and is shaded gray in these regions.}
\label{spectra1011}
\end{figure}

For the first run of measurements, we used a stack of \SI{10}{\per\cm} and \SI{11}{\per\cm} low-pass filters along with the magnetically loaded NDF at the \SI{4}{K} stage.
These are the standard filters typically used for optically coupling our cryostat to room temperature test equipment.
We found a large absorption feature at \SI{100}{GHz}, as shown in Figure~\ref{spectra1011}.
This sort of absorption feature is somewhat common for quasi-optical filter transmission well below the cutoff frequency.
We found that if we take the ratio of the spectra, we could recover a bandpass of the on-chip filter with a relatively flat transmission spectrum.
In order to test this spectral ratio technique, we needed to use another optical configuration.

\subsection{Measurement Run 2}

\begin{figure}[htbp]
\centerline{\includegraphics[width=\columnwidth]{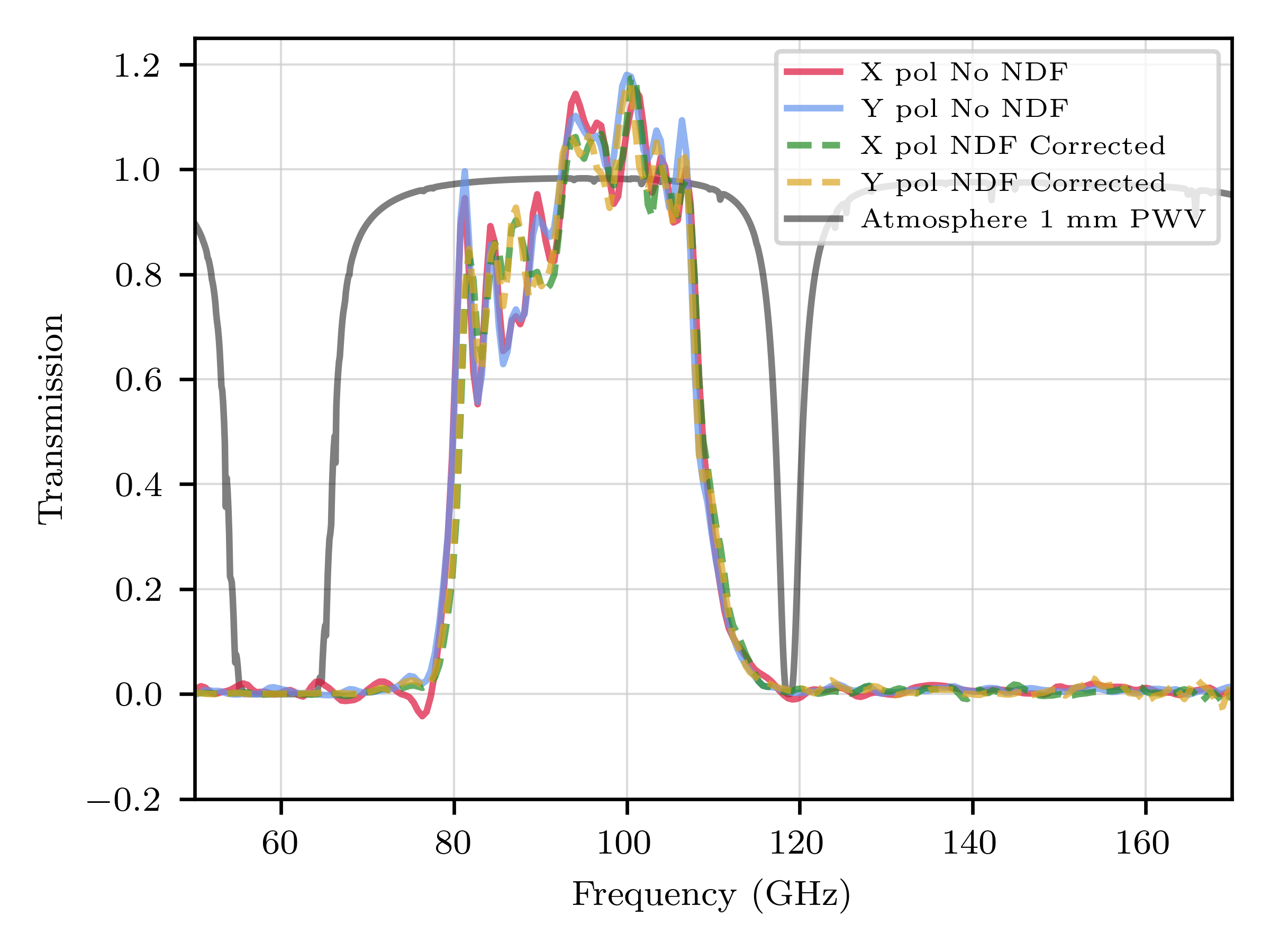}}
\caption{Frequency spectrum of the baseline pixel in both polarizations from measurement run 2, without the NDF, and from measurement run 3, with the NDF installed and spectral corrections applied.
Overplotted is the atmospheric transmission with \SI{1}{\mm} of precipitable water vapor~\cite{pardo2002atmospheric}.}
\label{baseline}
\end{figure}

In the second run, we replaced the \SI{10}{\per\cm} and \SI{11}{\per\cm} LPFs with a \SI{5.8}{\per\cm} LPF which has a more appropriate cutoff relative to the detector passband.
We replaced the NDF with a PTFE low-pass filter which should have less spectral dependence in absorption at these frequencies.
However, the pixel without bandpass filters was saturated with optical power and would not transition at the lowest achievable bath temperature, so we did not get data on this pixel split.
The baseline pixel data from measurement run 2, shown in Figure~\ref{baseline} (solid lines), does not have the absorption feature at \SI{100}{GHz} which can be attributed to changing the LPFs.

\subsection{Measurement Run 3}

\begin{figure}[htbp]
\centerline{\includegraphics[width=\columnwidth]{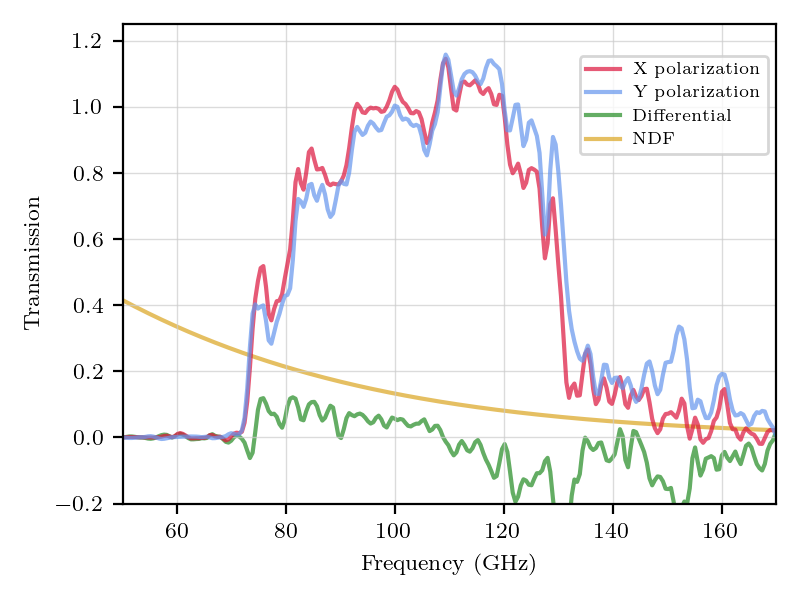}}
\caption{Measured frequency spectrum from the pixel design split without on-chip filters.
The transmission of the the X and Y polarizations has been corrected for the NDF transmission, shown in yellow.
}
\label{nobpf}
\end{figure}

Finally, in the last measurement run, we replaced the PTFE filter with the NDF to reduce optical loading and obtain spectra on both pixels.
The addition of the NDF reduces the optical power and allows the TESs to enter their superconducting transitions.
The bandpass measurements, corrected for the frequency-dependent attenuation of the NDF are shown in Figure~\ref{baseline} (dashed lines) for the baseline pixel with on-chip filters and in Figure~\ref{nobpf} for the pixel without on-chip filters.
Figure~\ref{onchip} shows the ratio of measured spectra of the baseline pixel with on-chip filters to that without on-chip filters to isolate the filter response. The simulated bandpass filter response is also shown for comparison.

\section{Discussion}

\begin{figure}[htbp]
\centerline{\includegraphics[width=\columnwidth]{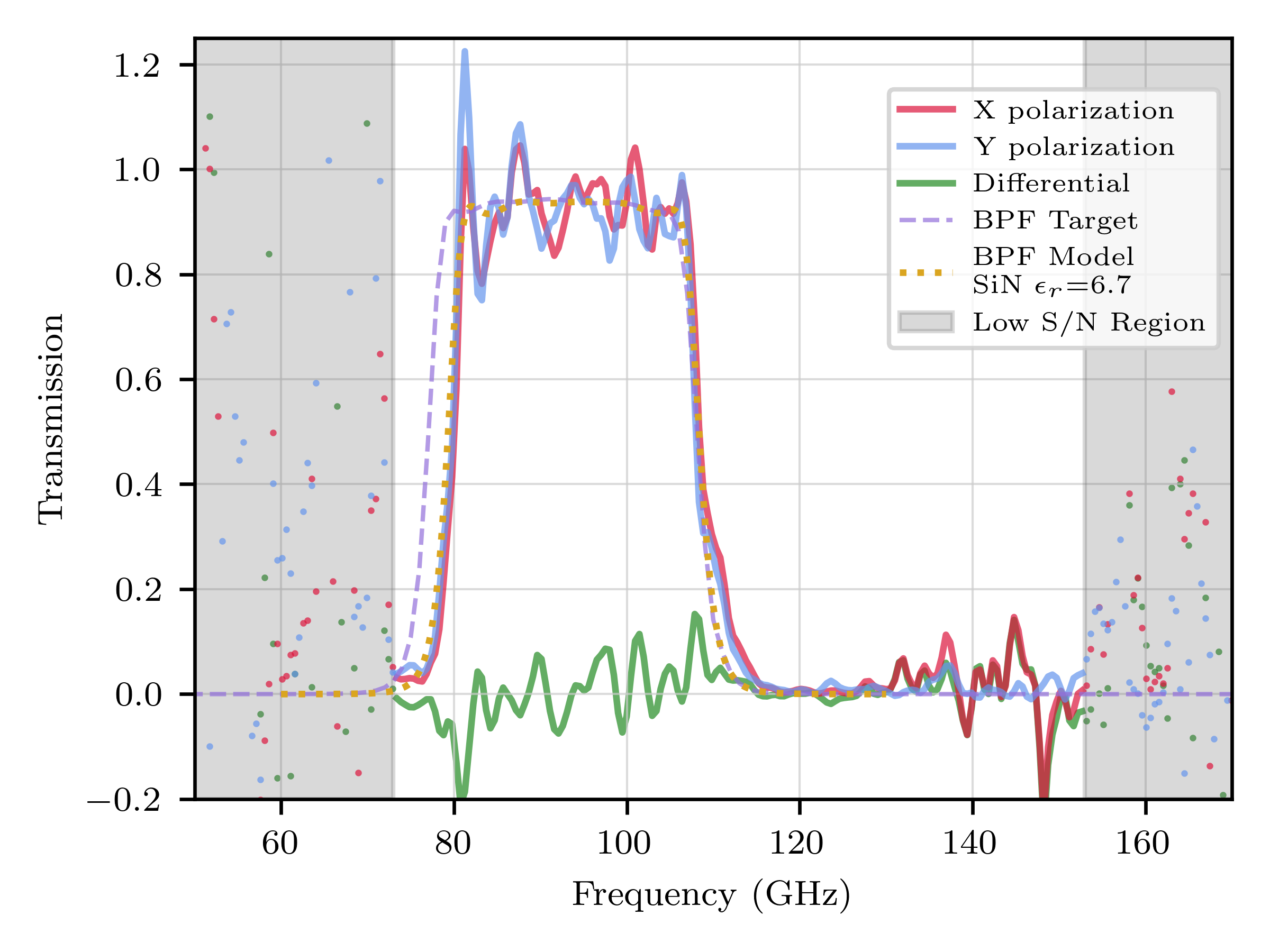}}
\caption{Measured and simulated frequency spectrum of the isolated on-chip filters.
This data is the ratio of the baseline pixel in Figure~\ref{baseline} measurement run 3 (dashed lines), and the pixel without on-chip filters in Figure~\ref{nobpf}.
The differential transmission between X and Y polarizations is minimal.
}
\label{onchip}
\end{figure}

We measured the spectral response of the CLASS pixel and found the $\mathbf{\text{\SI[bracket-negative-numbers = false]{-3}{dB}}}$ band edges to be \SI{80}{GHz} and \SI{108}{GHz}.
Figure~\ref{onchip} shows that the lower band edge is \SI{3}{GHz} higher than the target and the upper band edge is on target.
In a previous iteration of the fabrication, the band edges matched the design.
The lower band edge shift can be decomposed into an upward shift in frequency and a decrease in bandwidth.
One possible reason for the the entire band to shift is that the relative permittivity, $\epsilon_r$, of the SiN layer is not 6.8, as assumed, but a smaller value around 6.7, which is not an unreasonable fluctuation of the dielectric value.
The bandwidth of a stub filter is proportional to the ratio of the characteristic impedance of the line to the stub.
To reduce the bandwidth by \SI{3}{GHz}, there would need to be a significant over-etch of the microstrip transmission line width from \SI{5}{\um} to \SI{4.5}{\um}.
This would increase the impedance ratio of the transmission line to the stubs of the filter.
However, we expect the over-etch to be \SI{0.1}{\um} at most.
Another possibility is dielectric thickness variation; if the SiN was \SI{75}{nm} thicker it would shift the band upwards by 3~GHz.
However, this would be much thicker than expected.
These witness pixels were fabricated at the edge of a \SI{150}{mm} silicon wafer with the 37 pixel array in the center.
We've seen passband edges change with position on the wafer from up to \SI{+2}{GHz} at the center to \SI{-2}{GHz} at the edge in a \SI{90}{GHz} band~\cite{sierra2025simons}.
The exact mechanism for the band pass shift is still being explored.

Another notable characteristic in the baseline pixel spectrum, Figure~\ref{baseline}, is a step feature in the bandpass at \SI{92}{GHz} where the transmission is 28\% higher after the step.
We are able to conclude that this step does not come from the on-chip filter due to its absence in the isolated BPF spectrum in Figure~\ref{onchip}.

The spectral ratio technique is a useful tool for isolating other individual optical components.
Despite the limitations of the \SI{10}{\per\cm} and \SI{11}{\per\cm} filter configuration and the frequency-dependent absorption of the NDF, the on-chip filter response is still recoverable from the run~1 data and consistent with the results from run~3 shown in Figure~\ref{onchip}.

\begin{figure}[htbp]
\centerline{\includegraphics[width=\columnwidth]{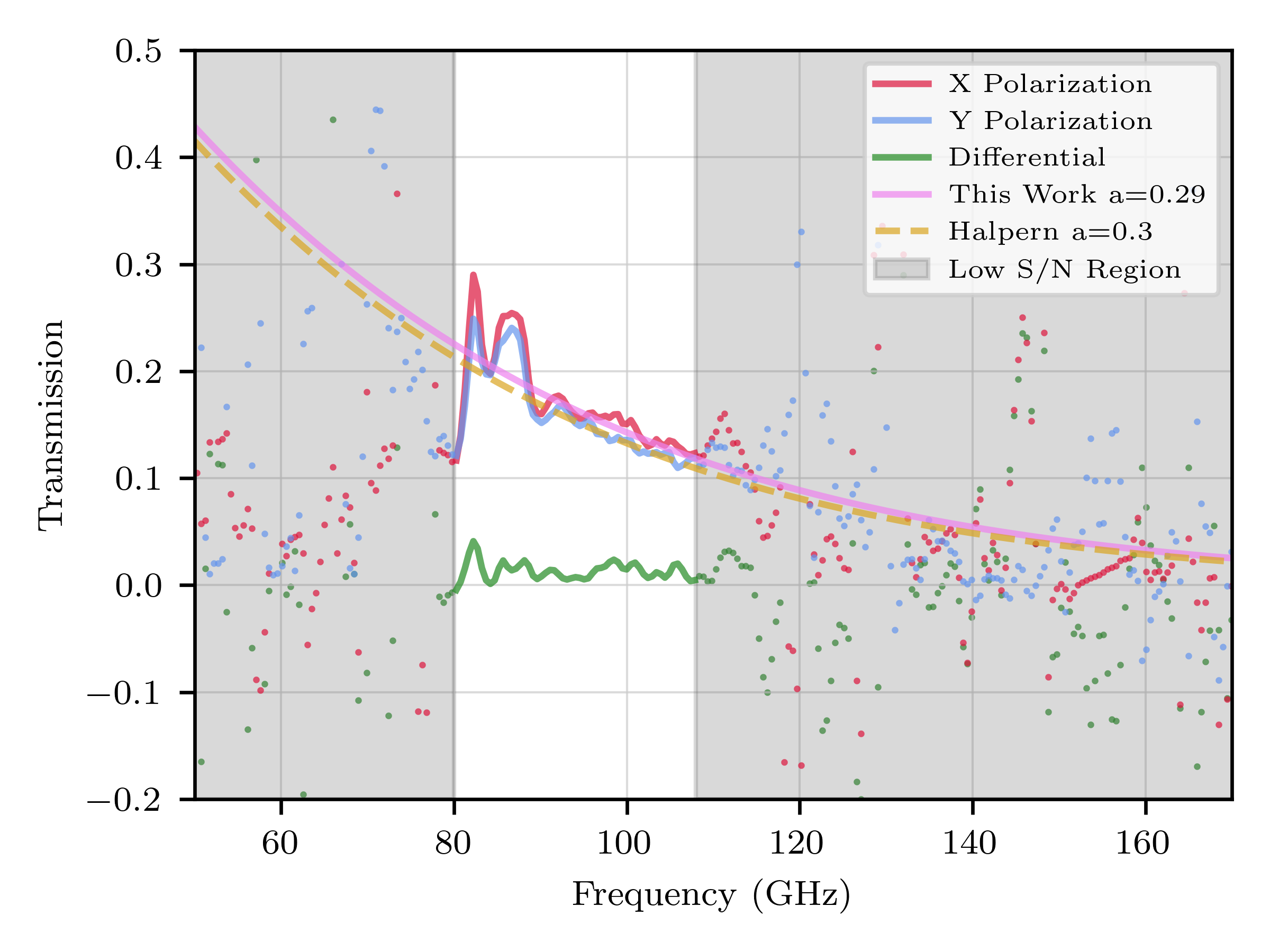}}
\caption{Transmission spectrum of the NDF made from
from the ratio of the run~2 and run~3 datasets of the baseline pixel in both polarizations, red and blue lines. The green line shows the differential transmission is consistent with zero. The best fit absorption model is shown in pink and can be compared to the literature value shown by the yellow dashed line.}
\label{ndf}
\end{figure}

We can also take the ratio of the spectra with and without the NDF of the baseline pixel in runs~2 and~3, shown in Figure~\ref{ndf}.
The PTFE filter, used in run 2, has a nearly flat transmission spectrum in this frequency range, so the resulting spectrum from the ratio is dominated by the response of the NDF~\cite{lamb1996miscellaneous}.
The attenuation coefficient is parameterized as $\alpha=a\nu^b$, where $\nu$ is frequency, and $a$ and $b$ are constants.
Since we are using the baseline pixel for this ratio, there is not enough broadband frequency coverage to constrain the frequency exponent, $b$, well, so we take the reference value of $1.2$~\cite{halpern1986far}. 
We find our best fit $a=0.29$, which is in good agreement with the reference value of $a=0.3$.
The signal-to-noise ratio is very low outside the band edges and is a limitation of the ratio method.
The differential bandpass of the NDF is minimal.

\begin{figure}[htbp]
\centerline{\includegraphics[width=\columnwidth]{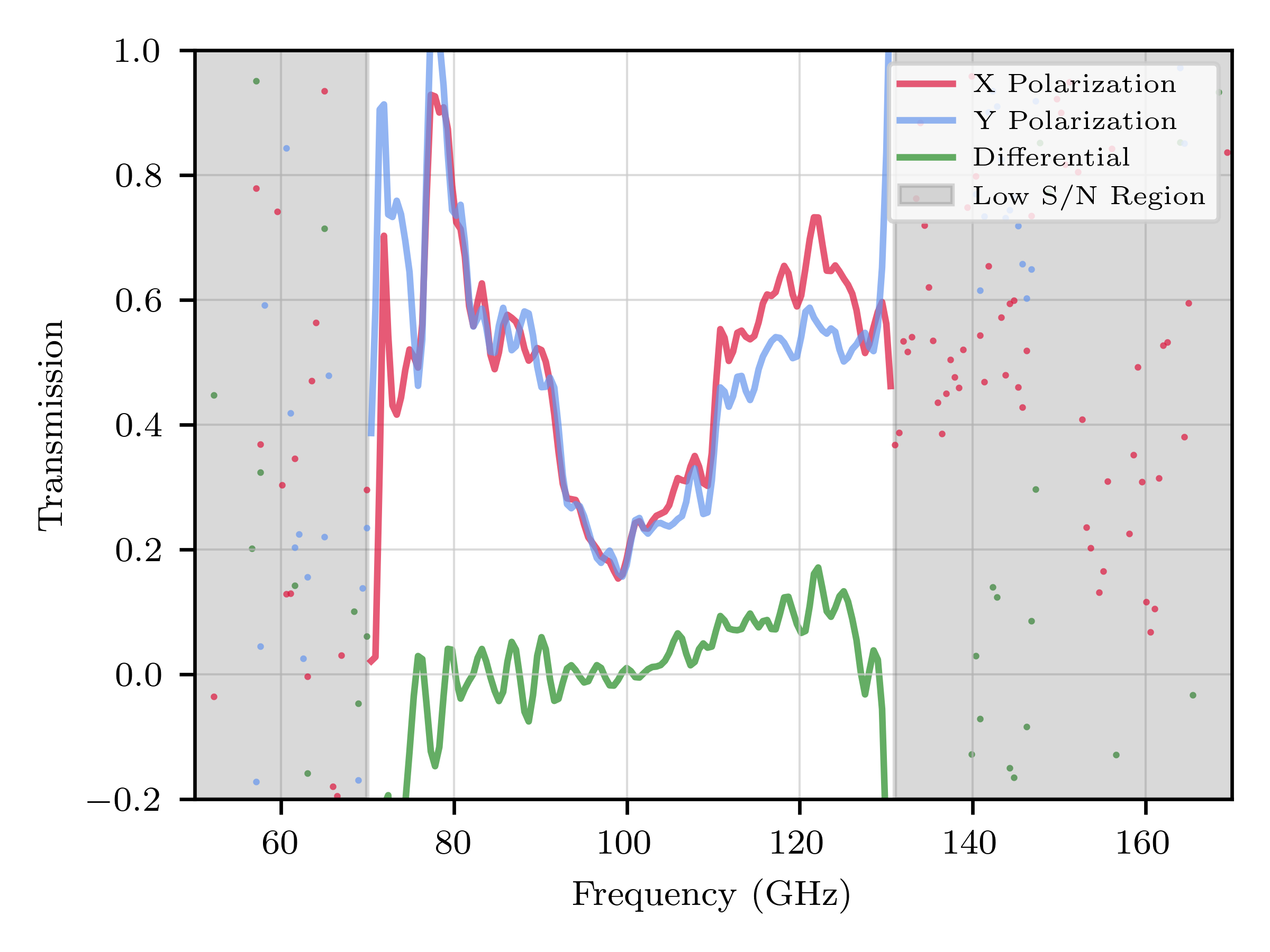}}
\caption{Measured differential transmission of the quasi-optical filters obtained from the ratio of the measurement run 1 (\SI{10}{\per\cm} and \SI{11}{\per\cm}) and measurement run 3 (\SI{5.8}{\per\cm}) datasets of the pixel split without on-chip filters.}
\label{dif_lpf}
\end{figure}

We can use the broader frequency range of the pixel without on-chip filters and the data from runs~1 and 3 to measure the differential transmission of the quasi-optical filters by taking their ratio.
The results provide a clear comparison of the two quasi-optical filter stacks, shown in Figure~\ref{dif_lpf}.
Although we cannot make an absolute measurement of the transmission of each filter with the ratio method, we can identify the impact on the overall system response.
The absorption feature in the run~1 data can be attributed to the \SI{10}{\per\cm} and \SI{11}{\per\cm} filters because it disappears when the filter is swapped out in later runs.

\section{Conclusion}

We measured the spectra of witness pixels from the CLASS \SI{90}{GHz} detector arrays and compared the baseline design to a design split without on-chip filters.
Including the design split without on-chip filters is useful for isolating the spectrum of the on-chip filters using the spectral ratio method.
We show that the bandpass of the on-chip filter is relatively flat and that a step feature at \SI{92}{GHz} in the baseline pixel response originates from outside the on-chip filters.
The same method of taking ratios of spectra can also be used to obtain the differential transmission of optical elements in the system, which we did for the quasi-optical low-pass filters and the neutral density filters.
These results provide valuable information for the calibration and analysis of the CLASS data, and the methods developed here will be applied to future characterization campaigns.

\section*{Notes}
Certain commercial products are identified to adequately specify the experimental study. This does not imply endorsement by NIST or that the instruments and/or products are the best available for the purpose.

\bibliographystyle{ieeetr}
\bibliography{lebib}

\vspace{12pt}

\end{document}